\documentclass[11pt,twoside]{article}

\usepackage{asp2004}
\usepackage{epsf}
\usepackage{psfig}
\usepackage{graphicx}

\begin{document}
\title{Stokes Inversion Techniques: Recent Advances and New Challenges} 
\author{L. R. Bellot Rubio}   
\affil{Instituto de Astrof\'{\i}sica de Andaluc\'{\i}a (CSIC), 
Apdo.\ 3004, E-18080 Granada, Spain}    

\begin{abstract} 
Inversion techniques (ITs) allow us to infer the magnetic, dyna\-mic,
and thermal properties of the solar atmosphere from polarization line
profiles. In recent years, major progress has come from the
application of ITs to state-of-the-art observations. This paper
summarizes the main results achieved both in the photosphere and in
the chromosphere. It also discusses the challenges facing ITs in the
near future.  Understanding the limitations of spectral lines,
implementing more complex atmospheric models, and devising efficient
strategies of data analysis for upcoming ground-based and space-borne
instruments, are among the most important issues that need to be
addressed. It is argued that proper interpretations of
diffraction-limited Stokes profiles will not be possible without
accounting for gradients of the atmospheric parameters along the line
of sight. The feasibility of determining gradients in real time from
space-borne observations is examined.
\end{abstract}

\section{Introduction}
We derive information on the physical properties of the solar
atmosphere by interpreting the polarization profiles of spectral
lines. Extracting this information is not easy, because the observed
profiles depend on the atmospheric parameters in a highly non-linear
manner through the absorption matrix and the source function vector.
To solve the problem, least-squares inversion techniques (ITs) based
on analytical or numerical solutions of the radiative transfer
equation were developed in the past. These methods compare the
observed Stokes profiles with synthetic profiles emerging from an
initial guess model atmosphere.  The misfit is used to modify the
atmospheric parameters until the synthetic profiles match the observed
ones. This yields a model atmosphere capable of explaining the
measurements, within the assumptions and limitations of the model.

The first ITs were proposed by \citet{b3 Ha72} and \citet{b3 Au77}.
Many other codes have been developed since then
\citep[cf.\ Table 1 in][]{b3 TI03}.
Today we have an IT for
almost any application we may be interested in: LTE or non-LTE line
formation, one-component or multi-component model atmospheres,
photospheric or chromospheric lines, etc. These codes have been used
intensively to study the magnetism of the solar atmosphere, and now
they are essential tools for the analysis of spectro-polarimetric
measurements.

This paper concentrates on recent achievements and future challenges 
of ITs. Additional information on ITs can be found in the 
reviews by del Toro Iniesta \& Ruiz Cobo (1996), Socas-Navarro 
(2001), and del Toro Iniesta (2003).

\section{Recent Advances}
There have been no significant improvements of classical least-squares
ITs in recent years: no new algorithms have appeared, and existing
codes have not been optimized for speed. However, the experience
accumulated has been used to develop new codes for specific purposes.
The complexity of both atmospheric and line-formation models has also
increased. For example, a few years ago the most sophisticated
inversions of Stokes profiles from sunspots were based on
one-component models with gradients of the physical parameters
(Westendorp Plaza et al.\ 2001), while two-component inversions are now performed 
on a routine basis \citep{b3 BR04b,b3 Bo04}.
Even more complex models are necessary to explain the net circular
polarization (NCP) of spectral lines emerging from sunspot penumbrae.
\citet{b3 SA05} used micro-structured magnetic atmospheres, assuming
that the penumbra is formed by optically thin magnetic fibrils.
\Citet{b3 Bo05} adopted an uncombed penumbral model with two different
components representing inclined flux tubes embedded in a more
vertical ambient field. Both models successfully reproduce the
anomalous Stokes $V$ profiles observed near the neutral line, for
visible and infrared (IR) lines considered separately. The synthetic
NCPs, however, are a bit smaller than the observed ones, implying that
there is still room for improvement.

An important advance in the field has been the development of
alternative methods for real-time analyses of large data sets. Other
significant achievements have come from the application of classical
ITs to state-of-the-art observations. These issues are examined in
more detail in the next subsections.

\subsection{Fast Inversion Codes}
Conceptually, the simplest inversion is one that uses a look-up table.
The idea is to create a database of synthetic Stokes profiles from a
large number of known model atmospheres, and look for the profile in
the database which is closest to the observed spectrum.  The
corresponding model is adopted as representative of the physical
conditions of the atmosphere from which the profile emerged. Despite
its simplicity, this method had seldom been put into practice until
\citet{b3 Re00} drew attention to Principal Component Analysis (PCA)
as a means to accelerate the search in the look-up table. By virtue of
PCA, the Stokes profiles can be expressed in terms of a few
coefficients only. The comparison between observed and synthetic
profiles is then performed very quickly, because the calculation does
not involve the many wavelength points describing the full line
profiles. PCA lies at the heart of several codes developed in the 
last years.

The database is the most critical component of any IT based on look-up
tables, as its size increases dramatically with the number of free
parameters. To keep this number to a minimum, only Milne-Eddington
(ME) atmospheres have been used for PCA inversions of photospheric
lines (e.g., Socas-Navarro, L\'opez Ariste, \& Lites 2001). Even under
ME conditions, the parameter space cannot be sampled very densely. The
discrete nature of the database introduces numerical errors, and so
PCA analyses are less accurate than least-squares ME inversions
\citep[e.g.,][]{b3 SL87}.
However, the method gives an idea of the quality of the fit in 
terms of the so-called PCA distance. When this distance is large, 
the observed profile cannot be associated with any profile in the
database, making it possible to identify pixels that deserve closer
attention.

A nice feature of PCA inversions is that the search algorithm is
independent of the database. The synthesis of Stokes profiles may be
very time-consuming, but the inversion will always be fast. This opens
the door to the analysis of lines for which atomic polarization
effects are important. \Citet{b3 LC02}
have developed
a PCA inversion code to exploit the diagnostic potential of the Hanle
effect in the \ion{He}{i} D$_3$ line at $587.6\,$nm.  The database is
created using a line formation code which solves the statistical
equilibrium of a He atom with 5 terms, in the presence of
magnetic fields. Coherences between fine-structure levels within each
atomic term are accounted for to treat the Zeeman and Hanle regimes,
including level crossing (incomplete Paschen-Back effect). This code
has been applied to prominences \citep{b3 LC03,b3 Ca03}
and spicules \citep{b3 LC05}.

The speed of PCA inversions makes it possible to handle large amounts
of data in real time. Since 2004, PCA is used at the French-Italian
THEMIS telescope to derive vector magnetic fields from MTR
measurements of the \ion{Fe}{i} $630\,$nm lines \citep{b3 LA06}. Full
maps are inverted in about 10\,min, which is more or less the time
needed to take the observations. At the telescope, PCA is very useful
for quick-look analyses, allowing one to select interesting targets or
to continue the observation of interesting regions. While real-time
analyses are appealing, it is important to keep in mind their
limitations. In most cases, a proper interpretation of the
observations will require more sophisticated ITs, which however can
use the results of PCA methods as initial guesses.

Another promising technique explored in recent years is Stokes
inversion based on artificial neural networks (ANNs). The idea was
introduced by Carroll \& Staude (2001) 
and developed by \citet{b3 SN03,b3 SN05a}.
Essentially, an ANN is an interpolation algorithm. One
starts by setting up the structure of the network, i.e., the number
of layers and the number of neurons in each layer. The input layer
receives the observations (the Stokes profiles or a suitable
combination thereof) and the last layer outputs the unknown
atmospheric parameters. The ANN must be trained before applying it 
to real data. To this end, Stokes profiles synthesized using
different ME atmospheres are presented to the ANN in order to find the
synaptic weights and biases of the neurons that return the model
parameters used to compute the profiles. The training process is
very slow but, once accomplished, the ANN will invert a full
map in a matter of seconds. Indeed, ANNs are the fastest ITs available
nowadays. For the moment, however, they have not been used in any
scientific application.

Several strategies have been explored to optimize the performance of
ANNs. It seems that the best choice is to invert one parameter at a time
with a dedicated ANN.  Finding all atmospheric parameters with a
single ANN is possible, but requires a larger number of neurons and
the training process becomes very complicated.

\citet{b3 SN05a} has shown that 
the mean field strengths inferred with the help of ANNs are reasonably
accurate from a statistical point of view. On average they resemble those
provided by ME inversions.  However, the errors can be very large for
individual pixels, as evidenced by the large r.m.s.\ uncertainties (Fig.~8 in
that paper indicates an uncertainty of 0.3\,kG for fields of $1.2\,$kG, i.e.,
a relative error of 25\%; the error is even larger for weaker
fields). This means that ANNs may be appropriate for quick-look analyses and
other applications where high precision is not required. For detailed studies
of physical processes, current ANNs seem not to be accurate enough.

A limitation of both PCA-based inversions and ANNs is the use of ME
atmospheres, which precludes the determination of gradients of
physical parameters from the observed profiles. It remains to be seen
whether these methods can be modified in order to reliably recover 
gradients along the line of sight (LOS). As discussed in Sect.~3.3, 
gradients appear to be essential for the analysis of observations at 
very high spatial resolution.

\subsection{Application of ITs to State-of-the-Art Observations}
Major breakthroughs have come from the application of ITs to
state-of-the-art observations, including spectro-polarimetry in the
near IR, simultaneous observations of visible and IR lines,
spectro-polarimetry of molecular lines, and observations at 
very high spatial resolution.

The development of polarimeters for the IR, most notably the Tenerife
Infrared Polarimeter \citep[TIP;][]{b3 MP99},
has opened a new window with lines that offer excellent magnetic
sensitivity and chromospheric coverage. One example is the \ion{He}{i}
triplet at $1083\,$nm, which has become an essential tool to
investigate the upper chromosphere. The formation of the triplet is
complex and not really well understood. However, since the triplet
lines are nearly optically thin, ME atmospheres provide a good
description of their shapes. An inversion code specifically
designed for \ion{He}{i} $1083\,$nm, HELIX, was presented by 
\citet{b3 La04}.
It is based on the Unno-Rachkovsky solution of the
radiative transfer equation and includes an empirical treatment of the
Hanle effect. Outside active regions, the linear polarization
profiles of \ion{He}{i} $1080.3\,$nm show the signatures of the Hanle
effect, which needs to be taken into account for correct retrievals of
vector magnetic fields \citep{b3 TB02}.
HELIX implements
the PIKAIA genetic algorithm, rather than the more common Marquardt
algorithm employed by other least-squares ITs.  Using the magnetic
information obtained with HELIX, \citet{b3 So03}
were able to trace individual coronal loops in an emerging flux
region. They found upflows at the apex of the loops and downdrafts near
the footpoints, which is what one expects for magnetic loops rising
from deeper layers. Other applications of the code have been presented
by \citet{b3 La05}, \citet{b3 OS05}, and \citet{b3 So06}.

Routine observations of the IR triplet of \ion{Ca}{ii} at $850\,$nm
are now possible with the MTR mode of THEMIS and the
Spectro-Polarimeter for INfrared and Optical Regions 
\citep[SPINOR;][]{b3 SN06a}
mounted on the Dunn Solar Telescope (DST) of the
National Solar Observatory at Sacramento Peak. The Ca IR 
triplet lines are excellent diagnostics of the chromosphere, with the
advantage that their interpretation is much simpler than that of other
chromospheric lines such as \ion{Ca}{ii} H, \ion{Ca}{ii} K, and
H$\alpha$. However, they still require non-LTE computations. 
Non-LTE inversions of the Ca IR triplet lines observed with
SPINOR have been presented by \citet{b3 SN05b} and \citet{b3 SN06b}.
Although these analyses are very demanding in terms
of computational resources, they hold great promise for quantitative
diagnostics of the thermal and magnetic structure of the solar
chromosphere.

\begin{figure}[!t]
\centering
\begin{minipage}[c]{0.495\textwidth}
\centering
\includegraphics[height=3.45cm]{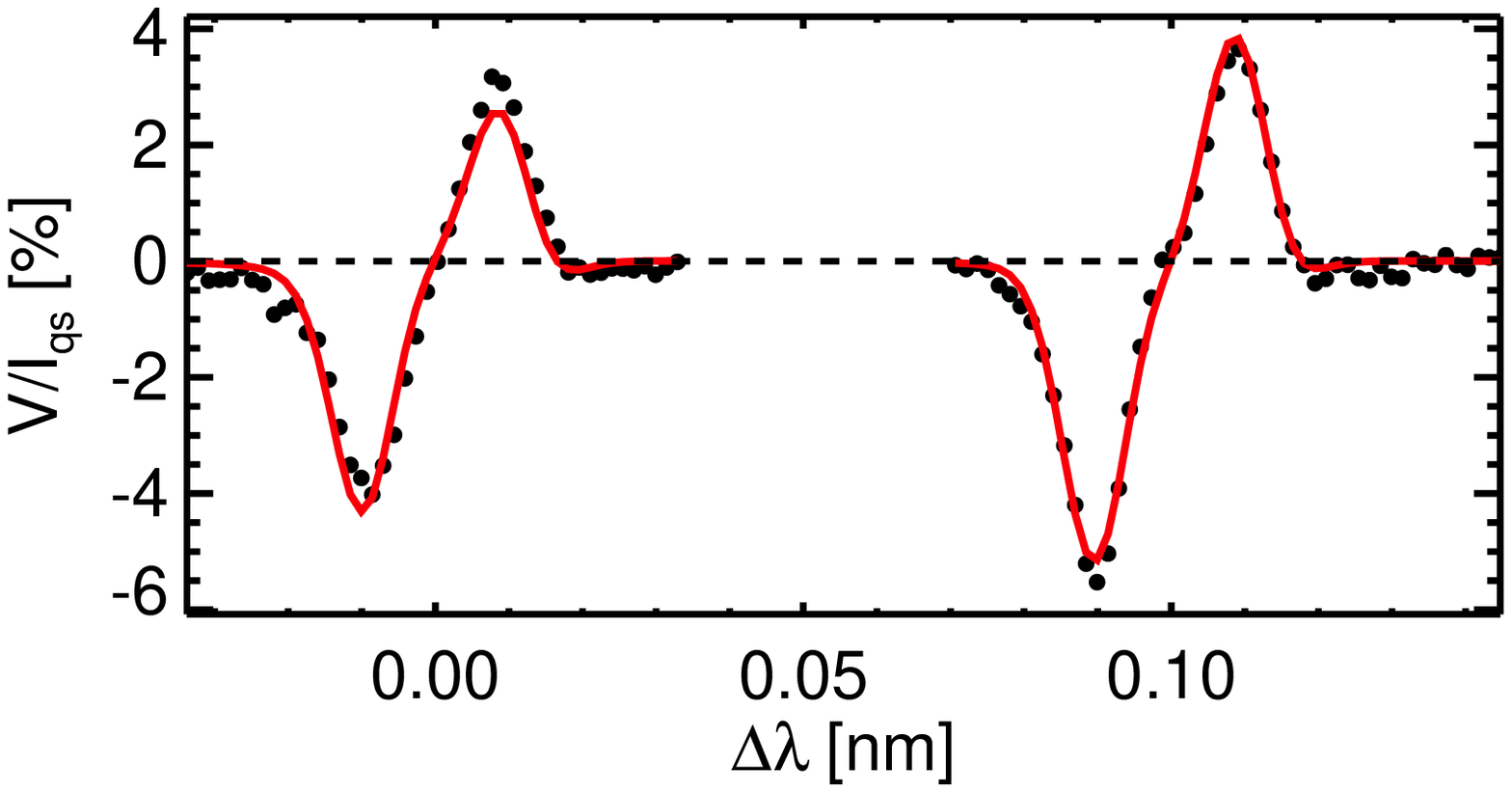}\vspace{8pt}
\centering
\includegraphics[height=2.7cm]{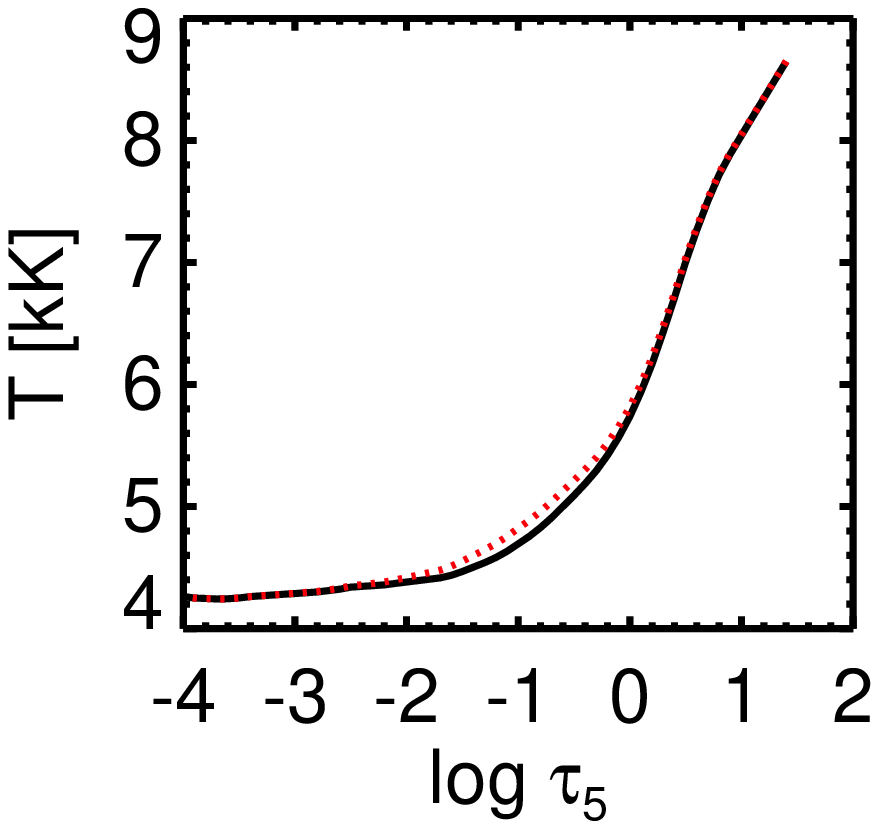}
\includegraphics[height=2.7cm]{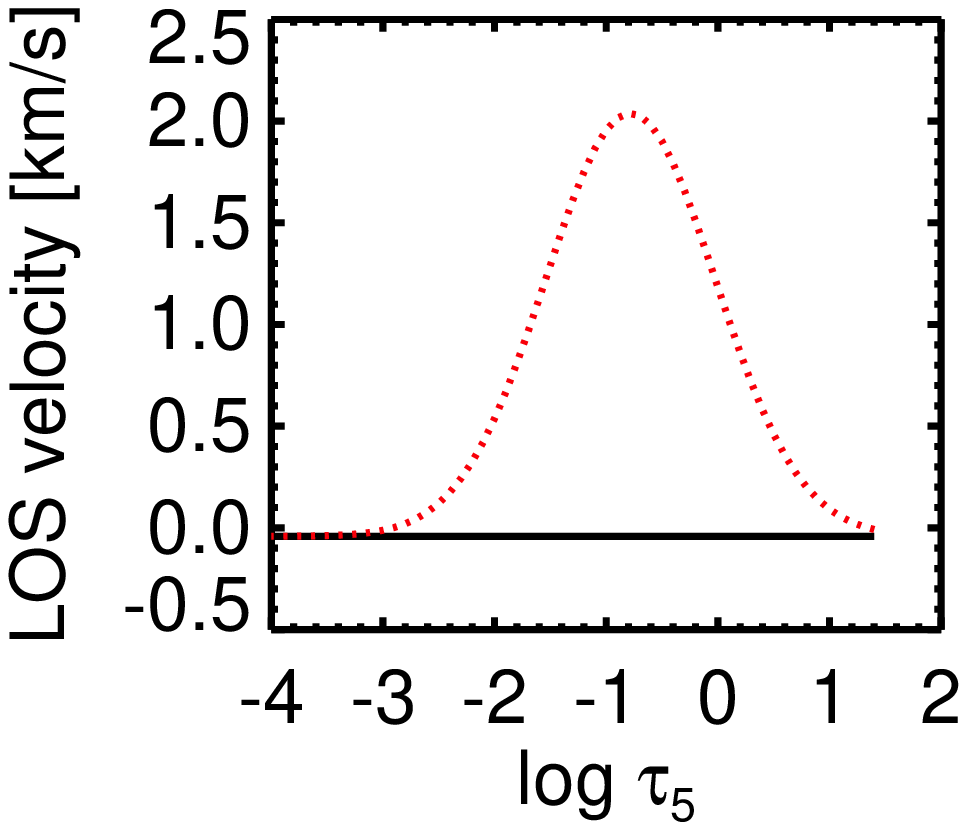}
\end{minipage}
\begin{minipage}[c]{0.495\textwidth}
\centering
\includegraphics[height=3.45cm]{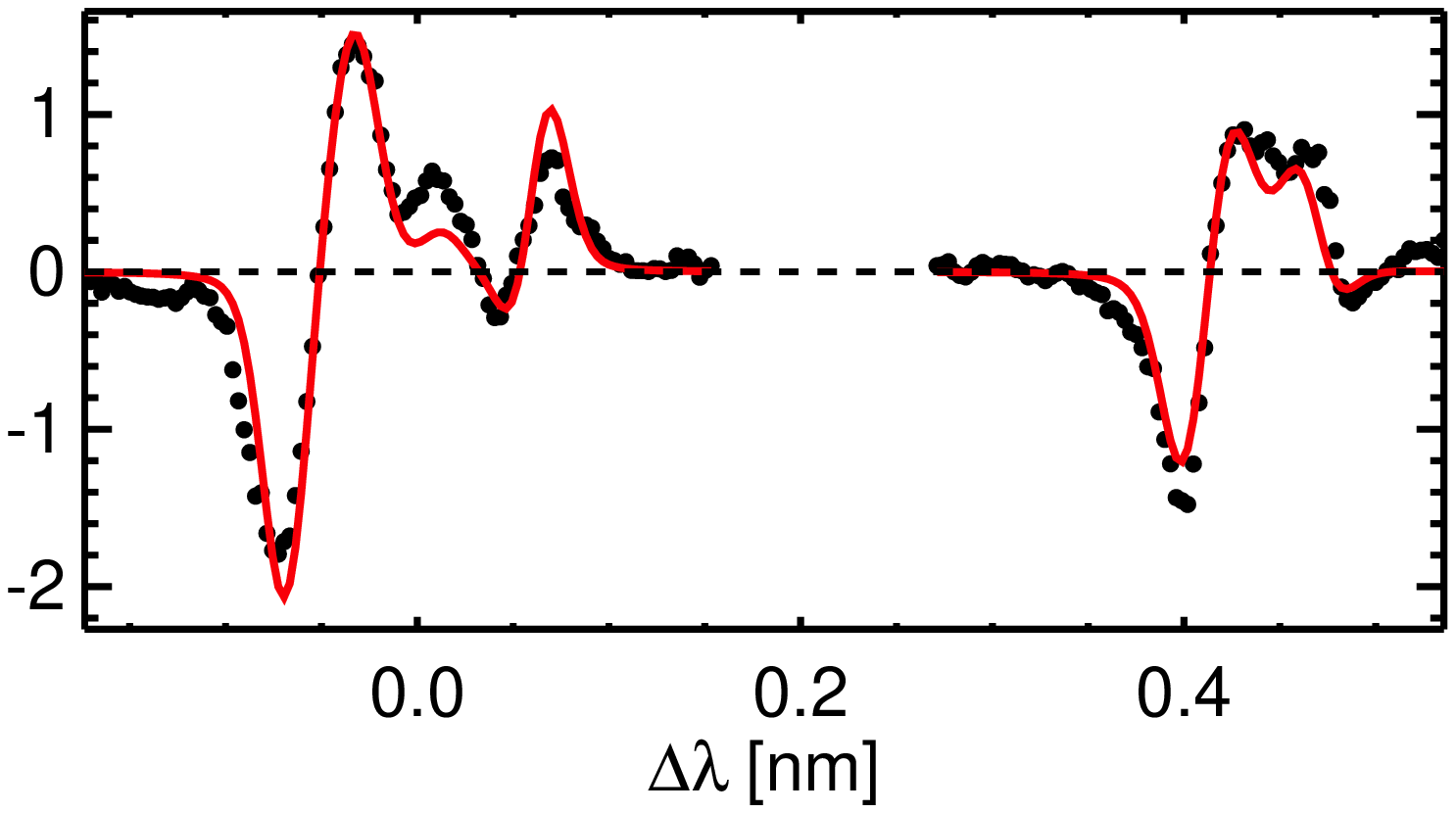}\vspace{8pt}
\centering
\includegraphics[height=2.7cm]{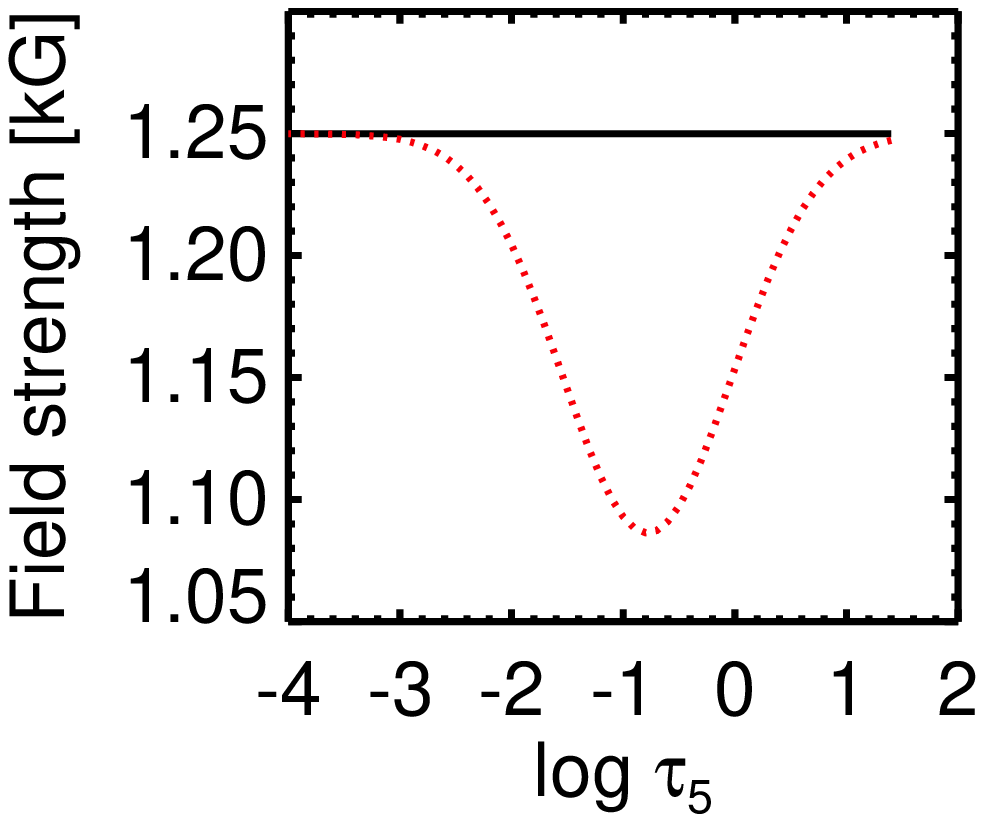}
\includegraphics[height=2.7cm]{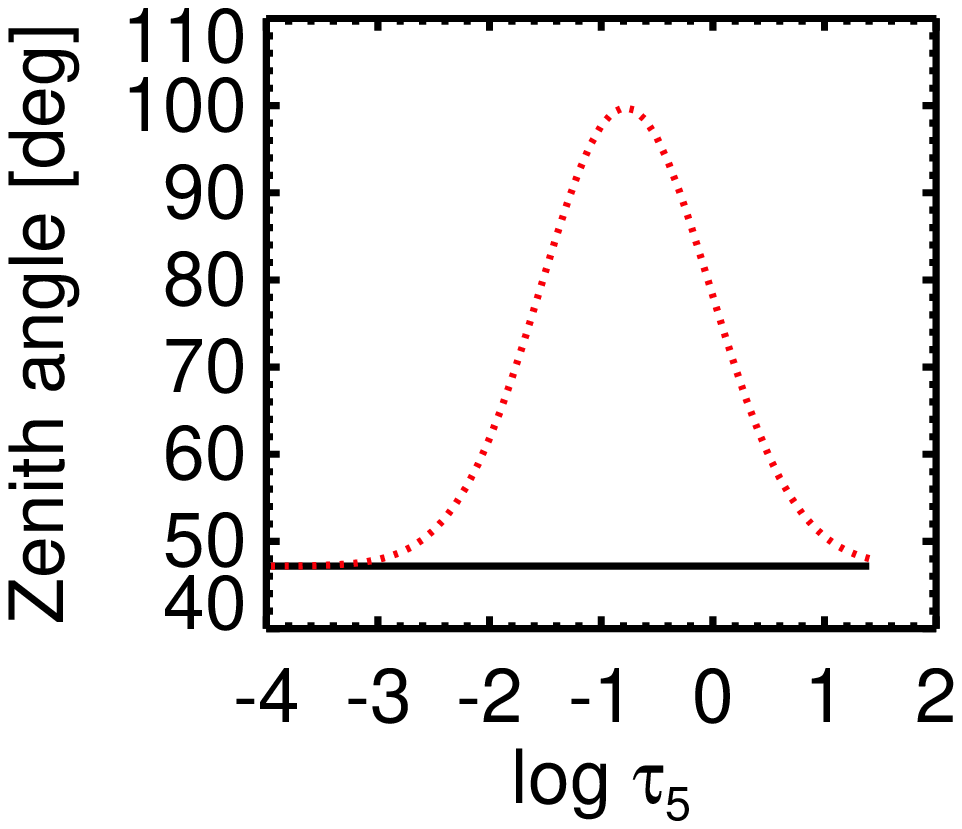}
\end{minipage}
\caption{{\em Top:} Stokes $V$ profiles of the pair of \ion{Fe}{i}
lines at 630 nm {\em (left)} and the \ion{Fe}{i} lines at 1565 nm {\em
(right)} observed simultaneously with POLIS and TIP in the limb-side
penumbra of AR\,10425, near the neutral line. Dots indicate the
observations, solid the best-fit profiles resulting from a
simultaneous inversion of the data. {\em Bottom:} Uncombed penumbral
model inferred from the inversion. From left to right: temperature,
LOS velocity, field strength, and field inclination in the background
(solid) and the flux-tube (dashed) components (from Beck, Bellot Rubio, 
\& Schlichenmaier, in preparation).
\label{fig1}}
\end{figure}  

Simultaneous observations of visible and IR lines improve the accuracy
of inversion results due to their different sensitivities to the
various atmospheric parameters \citep{b3 CS05}.
Several instruments are capable of such
observations, including TIP and POLIS 
\citep[POlarimetric LIttrow Spectrograph;][]{b3 Be05}
at the German VTT in Tenerife, and SPINOR at the DST.  Figure 1 shows
examples of Stokes $V$ profiles of the \ion{Fe}{i} lines at
$630.15\,$nm, $630.25\,$nm, $1564.8\,$nm, and $1565.2\,$nm observed in
the penumbra of AR\,10425 near the neutral line. These profiles were
taken strictly simultaneously with TIP and POLIS on August 9,
2003. Both the visible and IR lines exhibit large asymmetries and even
pathological shapes, suggesting the presence of two magnetic
components in the resolution element. We have carried out an analysis
of these profiles in terms of an uncombed penumbral model using the
code described by \citet{b3 BR03}.
The best-fit profiles from the simultaneous inversion are represented
by the solid lines. The quality of the fits is certainly remarkable,
but it should be stressed that even better fits are achieved from the
inversion of only the visible or the IR lines. Indeed, fitting both
sets of lines simultaneously is much more difficult, because a model
atmosphere appropriate for the visible lines may not be appropriate
for the IR lines, and vice versa. Measurements in two or more spectral
regions thus constrain the range of acceptable solutions.  The lower
panels of Fig.~1 show the uncombed model resulting from the inversion
(Beck, Bellot Rubio, \& Schlichenmaier, in preparation). The dashed
lines represent a penumbral flux tube with larger LOS velocities than
the background atmosphere, indicated by the solid lines. The field is
weaker and more horizontal in the tube.  These inversions confirm that
the uncombed model is able to explain the observed shapes of both
visible and IR lines.  At the same time, they allow us to determine
the position and width of the penumbral tubes, which is not easy from
visible or IR lines alone. Other examples of the analysis of
simultaneous measurements in the visible and IR are given by \citet{b3
BB05} and \citet{b3 Be06}.

The new polarimeters have also extended our capabilities to observe
molecular lines, providing increased thermal sensitivity. Molecular
lines are mostly seen in sunspot umbrae, because higher temperatures
dissociate the parent molecules. Usually, they show smaller
polarization signals than atomic lines. Two codes capable of inverting
molecular lines are SPINOR \citep{b3 FS98}
and the one developed by \citet{b3 AR04}.
SPINOR has been used to analyze 
the OH lines at $1565.2\,$nm and $1565.3\,$nm, improving the
determination of umbral temperatures \citep{b3 Ma03}.
The second code has been employed to invert the CN lines at $1542\,$nm
\citep{b3 AR05}.
These lines are
very interesting because in the umbra they show large linear
polarization signals but very small Stokes $V$ signals, just the
opposite behavior of atomic lines.

Finally, major progress has come from the application of ITs to high
spatial resolution spectroscopic and spectro-polarimetric
measurements. The main advantage of high spatial resolution is that
the results are less dependent on filling factor issues.  An example
of the inversion of high spatial resolution Stokes profiles is the
work of \citet{b3 SN04},
who derived the magnetic and
thermal properties of umbral dots from observations taken with the La
Palma Stokes Polarimeter \citep{b3 MP99}
at a resolution of about 0\farcs7. Another promising type of measurements
are those provided by Fabry-P\'erot interferometers such as the
Interferometric BIdimensional Spectrometer \citep[IBIS;][]{b3 Ca06}
and the TElecentric SOlar Spectrometer \citep[TESOS;][]{b3 Ke98},
which has recently been equipped with the KIS/IAA Visible
Imaging Polarimeter \citep[VIP;][]{b3 BR06b}.
Combined with adaptive optics systems, these instruments
perform 2D vector spectro-polarimetry at high spatial, spectral, and
temporal resolutions, which is necessary to investigate fast
processes in large fields of view. An example of the inversion of 2D
spectroscopic measurements with TESOS is the derivation of the thermal
and kinematic properties of a sunspot penumbra at different heights in
the atmosphere (\citealt{b3 BR06}; see also \citealt{b3 BR04a}).
The angular resolution of these observations is 0\farcs5.

In the future, significant progress may come from the routine
inversion of lines showing hyperfine structure, such as \ion{Mn}{i}
$553.77\,$nm and $874.09\,$nm (L\'opez Ariste, Tomczyk, \& Casini 2002).  
These lines exhibit sign reversals in the core of Stokes $V$ and
multiple peaks in Stokes $Q$ and $U$ for weak fields. Interestingly,
the shape of the anomalies depends on the magnetic field strength,
rather than on the magnetic flux. Such an unusual behavior can be used
to investigate the magnetism of the quiet Sun. In fact, from the shape
of the observed profiles it would be possible to determine directly
the strength of the magnetic field, even in the weak field regime.  To
exploit the diagnostic potential of these lines, however, it is
necessary to implement the appropriate Zeeman patterns in existing ITs
and to lower the noise level of current observations, which is barely
enough to detect the subtle signatures induced by hyperfine structure.

\section{Challenges}
ITs have proven to be essential tools to characterize the properties 
of the solar atmosphere. Their application to high precision
spectro-polarimetric measurements, however, has started to raise
concerns on the limitations of some spectral lines. This is an
important problem that deserves further investigation. Other
challenges facing ITs in the near future include the implementation of
more realistic model atmospheres and the development of strategies for
the analysis of the large amounts of data to be delivered by upcoming
instruments.

\subsection{Limitations of Spectral Lines}
\citet{b3 MG06}
have shown that it is possible to fit a given set of Stokes
profiles of the pair of \ion{Fe}{i} lines at $630\,$nm with very
different field strengths by slightly changing the temperature
stratification and the microturbulent velocity, for typical conditions
of quiet Sun internetwork regions. The reason is the different
formation height of the two lines. This quite unexpected result
suggests that one cannot determine reliable internetwork
field strengths from \ion{Fe}{i} $630.15\,$nm and $630.25\,$nm 
without a prior knowledge of the actual temperature stratification. 
Other lines such as \ion{Fe}{i} $524.71\,$nm and \ion{Fe}{i} 
$525.02\,$nm could provide the necessary information.

We have found a similar problem with the \ion{Fe}{i} $630\,$nm lines
even in the umbra, where the strong field regime applies. More
specifically, we have detected a cross-talk problem between the stray
light coefficient, the temperature, and the magnetic field strength
and inclination \citep[see][]{b3 CS06}.
The results of one-component inversions of umbral profiles with the
stray light contamination as a free parameter do differ from those in
which the stray light factor is fixed to the value inferred from a
simultaneous inversion of visible and IR lines.  With a mere
difference of 7\% in the stray light coefficient, the temperatures at
$\tau_5=1$ from the two inversions may differ by up to $150\,$K, and
the field strength by about $200\,$G. The fits are equally good in
both cases, so it is not possible to decide which inversion is
better. It appears that the relatively small Zeeman splitting of the
\ion{Fe}{i} $630\,$nm lines does not allow to clearly distinguish
between larger stray light factors and weaker fields, which produces
cross-talk among the various atmospheric parameters.

Given these concerns, more detailed studies of the limitations of
visible and IR lines seem warranted, for a better understanding of the
results obtained from them. To minimize the risk of cross-talk
problems in the inversion, it is desirable to use simultaneous
measurements in different spectral ranges. This will require
modifications of current inversion codes to account for different
stray light levels and different instrumental profiles in the
different spectral ranges.

\subsection{Implementation of More Realistic Atmospheric Models}
The new observational capabilities, in particular the availability of
simultaneous observations of visible and IR lines, offer us a
unique opportunity to increase the realism of the atmospheric models
implemented in existing ITs. The need for better models is indicated
by the small (sometimes systematic) residuals observed in inversions
of profiles emerging from complex magnetic structures.

As an example, consider the uncombed penumbral model. Right now we use
two different lines of sight to represent the background and flux-tube
atmospheres (cf.\ \citealt{b3 BR03,b3 Bo05}),
but this is a very simplistic approximation. The ambient
field lines have to wrap around the flux tube, hence the properties
of the background cannot be the same far from the tube and close to
it. This may have important consequences for the generation of
asymmetrical Stokes profiles. In much the same way, the flux tube is
not always at the same height within the resolution element, because
the magnetic field is not exactly horizontal. Therefore, different
rays will find the tube at different heights. Finally, lines of sight
crossing the center of the tube sense the properties of the tube over
a larger optical-depth range than lines of sight crossing the tube at
a distance from its axis. Neither of these effects are modeled by
current inversion codes. Probably, the subtle differences between
observed and best-fit profiles (cf.\ Fig.~\ref{fig1}) would disappear
with a more complex treatment of the magnetic topology of sunspot 
penumbrae.

\subsection{Analysis of Data from Next-Generation Instruments}
\label{high_resolution}
Stokes polarimetry at the diffraction limit is needed to study the
physical processes occurring in the solar atmosphere at their
intrinsic spatial scales. We are pushing our technological
capabilities to the limit by building grating spectro-polarimeters and
filter magnetographs for diffraction-limited observations.  On the
ground, examples of already operational or upcoming state-of-the-art
instruments include TIP, POLIS, DLSP, SPINOR, IBIS and TESOS+VIP. Among
space-borne instruments we have the spectro-polarimeter and filter
polarimeter onboard Solar-B, IMaX onboard SUNRISE, HMI
onboard SDO, and VIM onboard Solar Orbiter.

These instruments will deliver data of unprecedented quality in terms
of spatial and spectral resolution. We hope to further our
understanding of the solar magnetism with them. However, the success
of this endeavor will critically depend on our ability to extract in
an appropriate way the information contained in the observations. We
do not only want to investigate the morphology and temporal evolution
of the various solar structures from diffraction-limited images, but
also to derive their magnetic and kinematic properties accurately
using polarization measurements. Reliable determinations of vector
magnetic fields call for least-squares inversions. The problem is 
the enormous data flows expected: classical least-squares ITs are
considered to be too slow for real-time analyses of the observations.
This is the reason why it is taken for granted that the most
sophisticated inversions of the data will be based on ME models. 
The question naturally arises as to whether or not ME atmospheres are
appropriate for the interpretation of Stokes measurements at very
high spatial resolution.

\begin{figure}[!t]
\centering
\includegraphics[scale=.36]{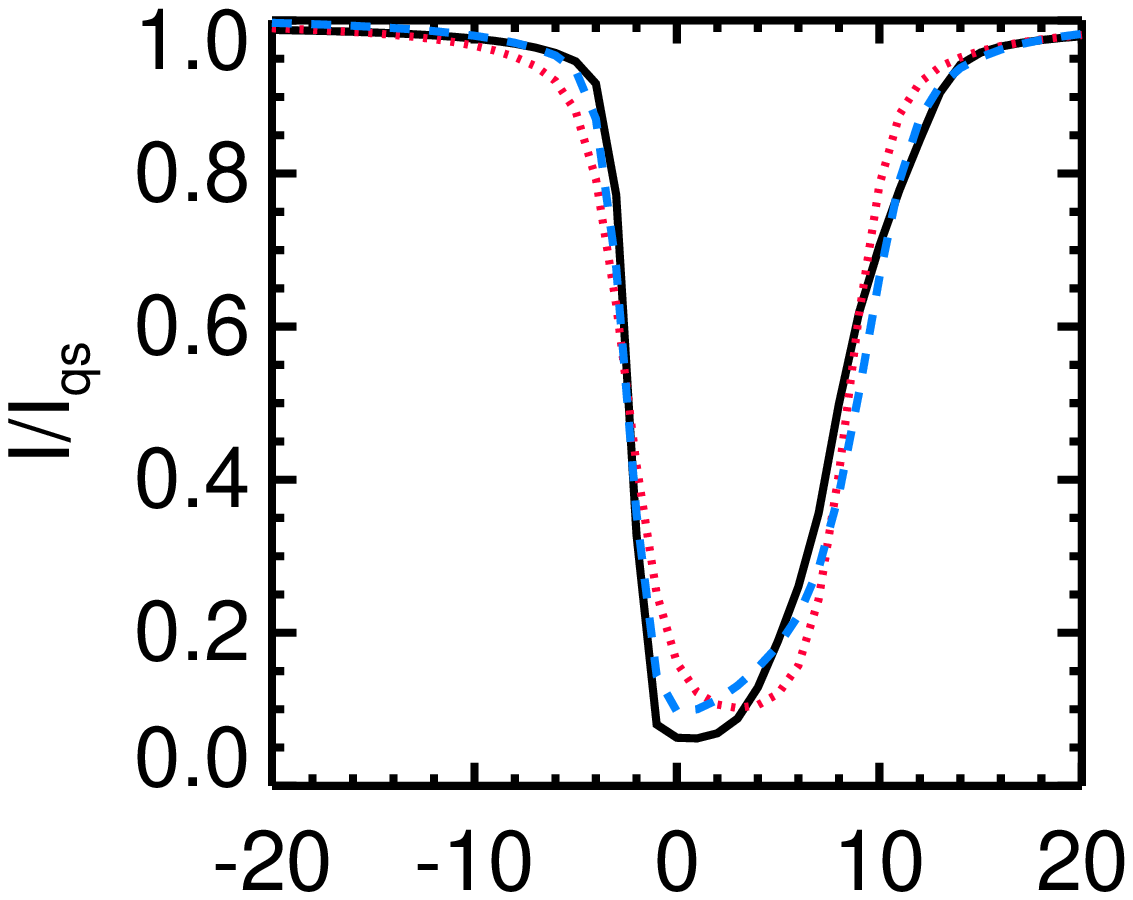}
\includegraphics[scale=.36]{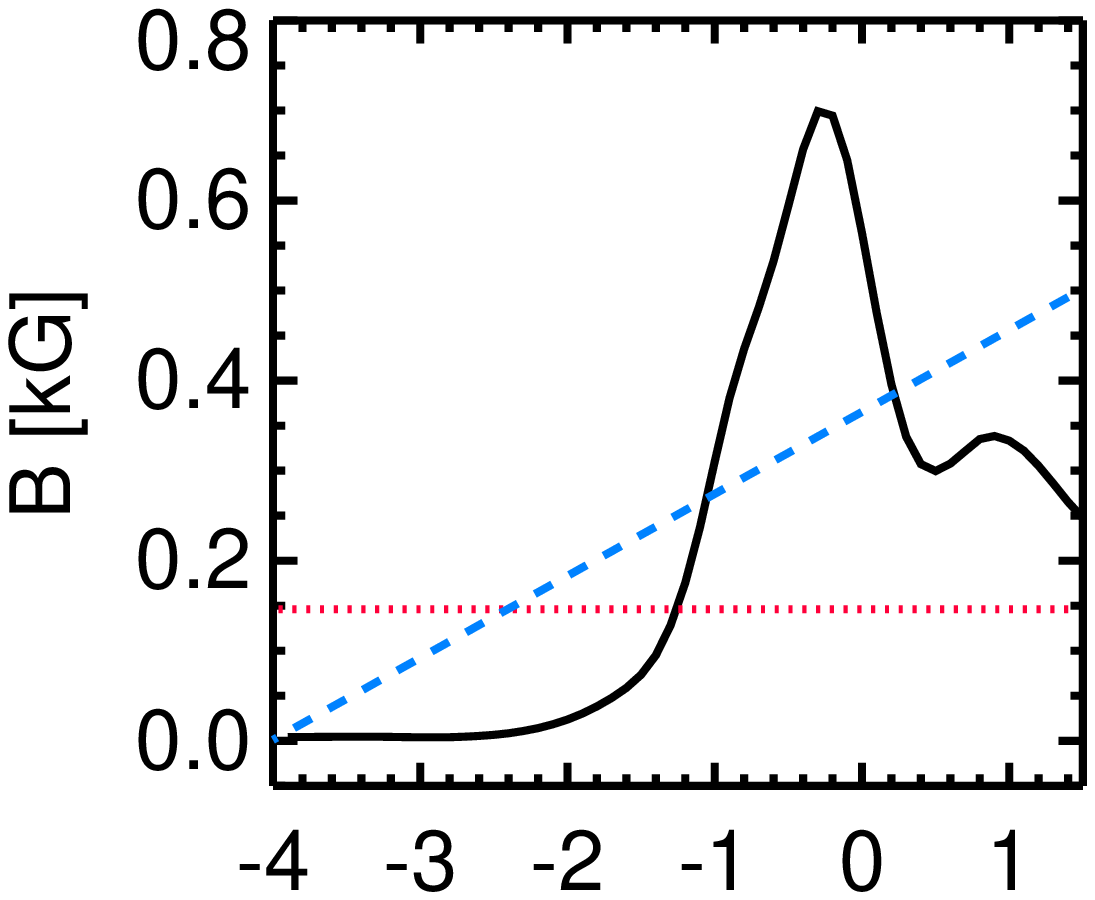}
\includegraphics[scale=.36]{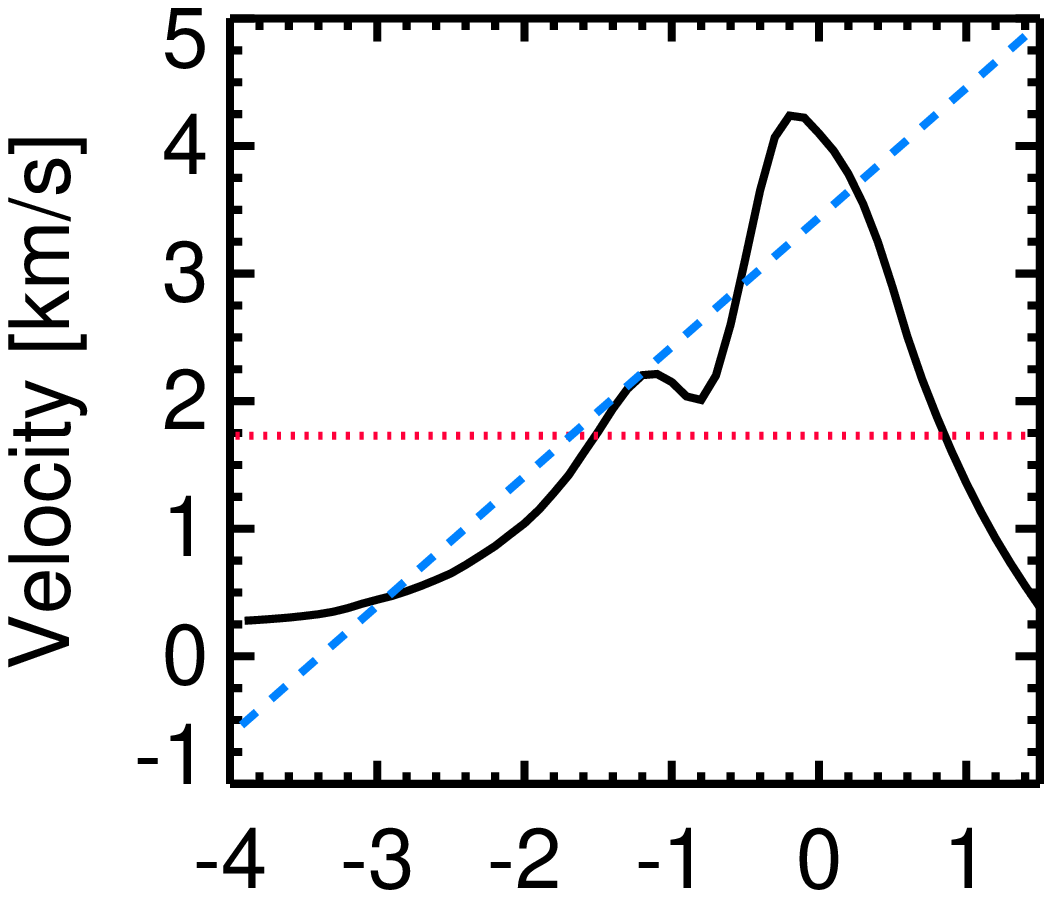}\vspace{1pt}
\centering
\includegraphics[scale=.36]{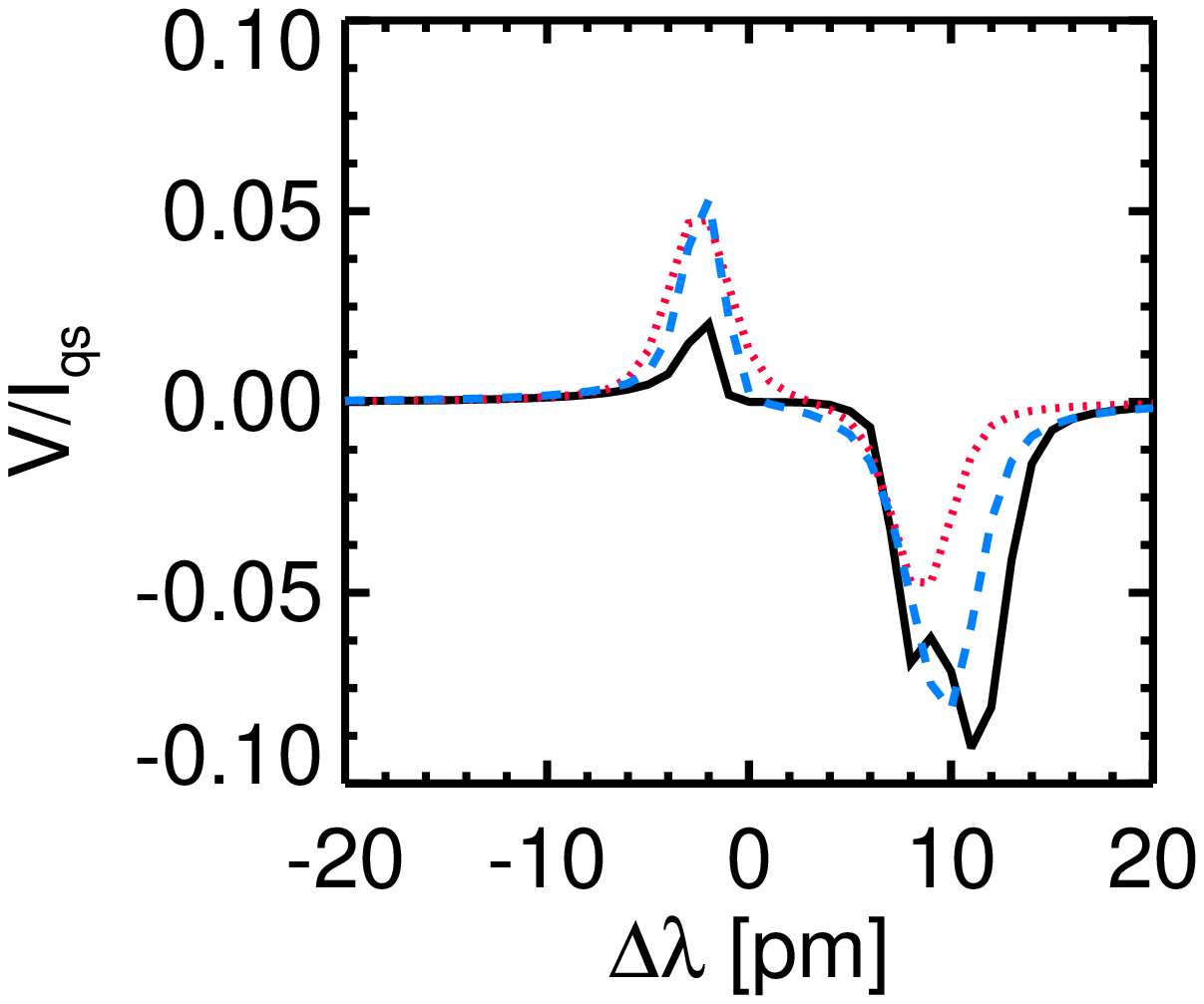}
\includegraphics[scale=.36]{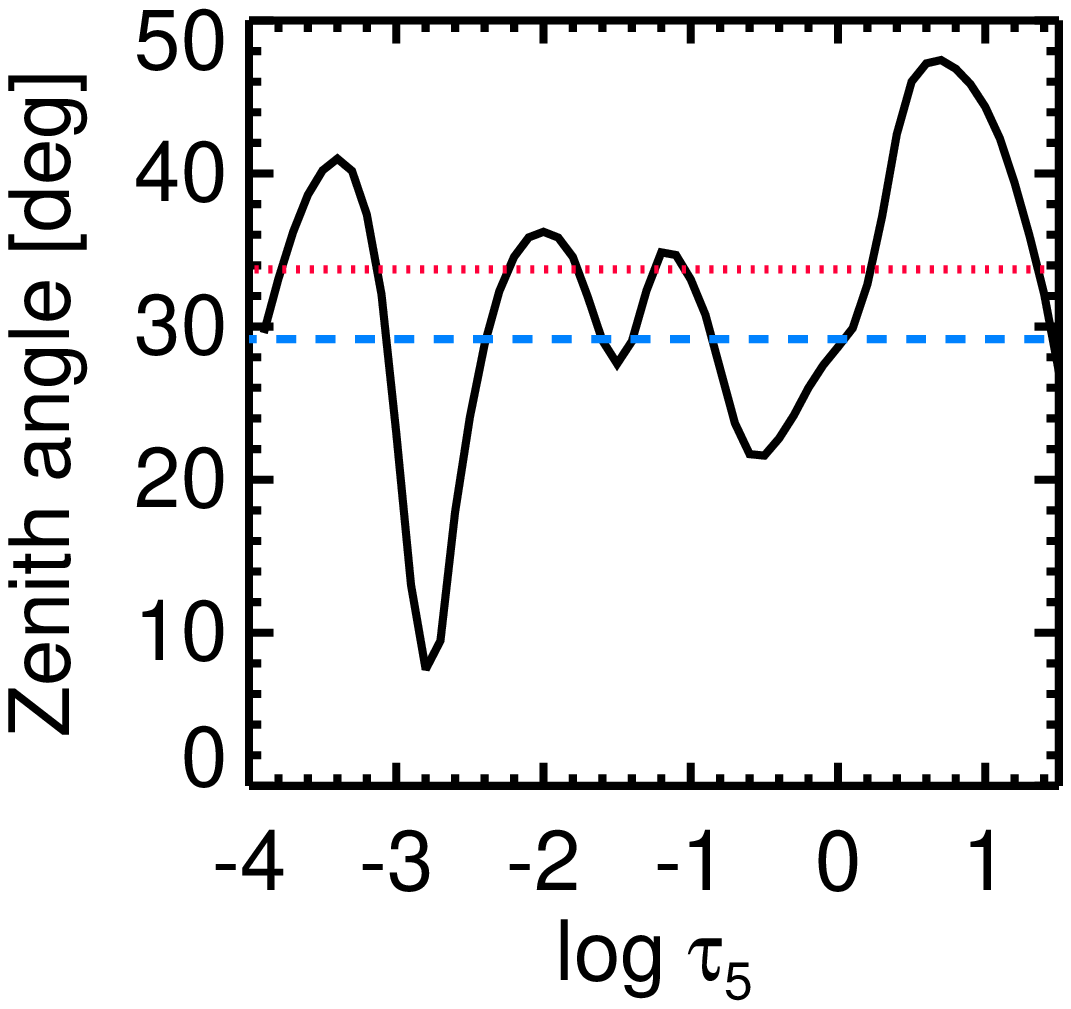}
\includegraphics[scale=.36]{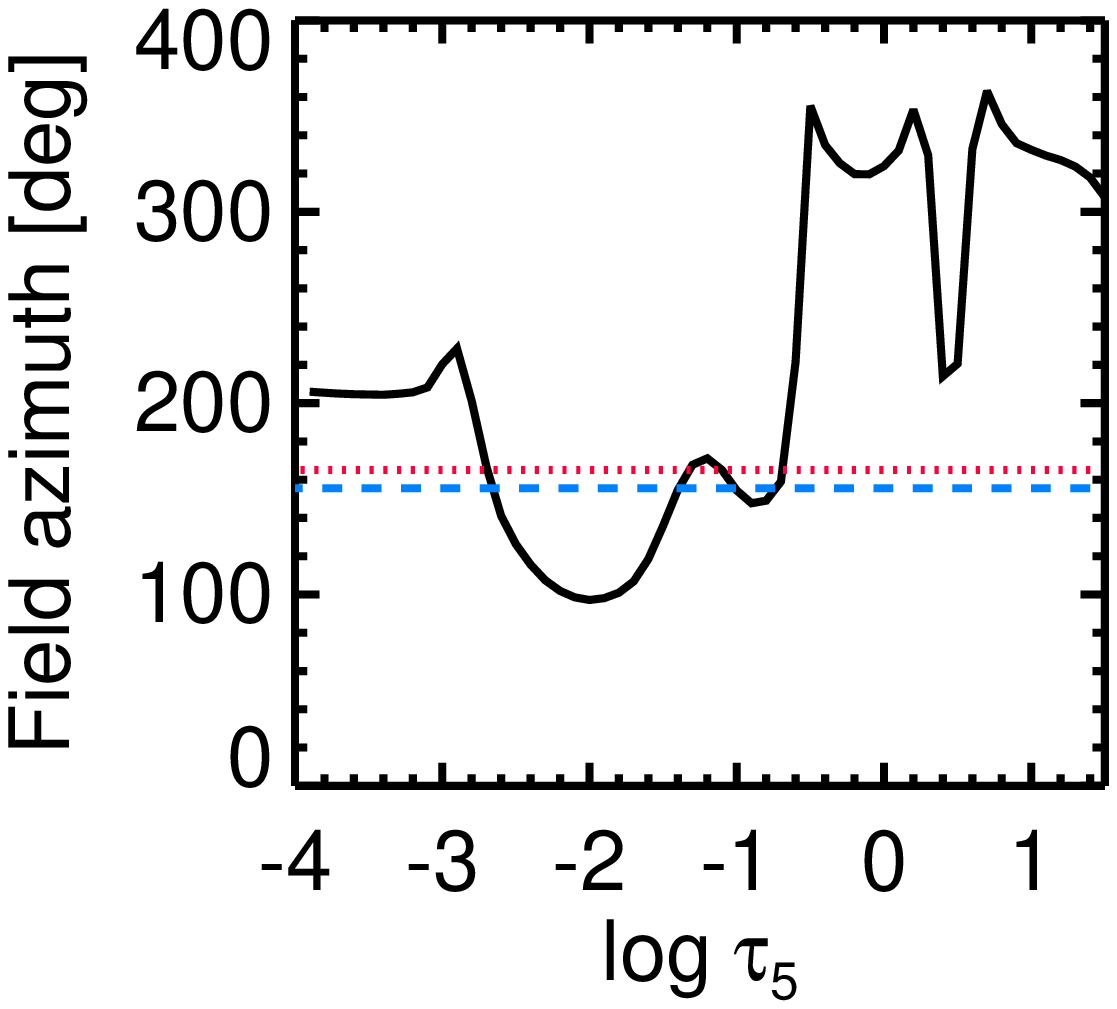}
\caption{{\em Left:} High spatial resolution Stokes $I$ and $V$
profiles of \ion{Fe}{i} 525.06~nm emerging from an intergranular lane
as computed from MHD simulations (solid). Dotted and dashed lines
represent the best-fit profiles from a ME inversion and a SIR
inversion with gradients, respectively. {\em Middle and right:}
Stratifications of atmospheric parameters used for the spectral
synthesis (solid). The results of the ME inversion and the SIR
inversion are given by the dotted and dashed lines, respectively.
\label{fig2}}
\end{figure}  

To shed some light on this issue, I have used the MHD simulations of
\citet{b3 Vo05}
to synthesize the Stokes profiles of the IMaX
line (\ion{Fe}{i} $525.06\,$nm) emerging from a typical magnetic
concentration in an intergranular lane. The atmospheric parameters
needed for the calculation have been taken from a simulation run 
with average magnetic flux density of $140\,$G. Figure~\ref{fig2} displays
the atmospheric stratifications and the corresponding Stokes profiles
at 0\farcs1 resolution (solid lines). The dotted lines show the
results of a ME inversion of the synthetic profiles.  As can be seen,
the fits to Stokes $I$ and $V$ are not very successful, due to the
extreme asymmetries of the profiles. The atmospheric parameters
inferred from the ME inversion are some kind of average of the real
stratifications, but the significance and usefulness of these average
values are questionable when the parameters feature such strong
variations along the LOS. An analysis of the same profiles with SIR
\citep{b3 RT92}
allowing for vertical gradients of field strength and velocity yields
much better fits to Stokes $I$ and $V$ (dashed lines). Although the
fits can still be improved, the important point is that this simple
SIR inversion is able to recover the gradients of field strength and
velocity with fewer free parameters than the ME inversion (8 as opposed
to 9). This additional information could be essential to understand
many physical processes, so it is important to have it.

Present-day computing resources are sufficient to determine gradients
from the high spatial resolution observations delivered by {\em
grating instruments} such as POLIS and the Solar-B spectro-polarimeter
(Solar-B/SP; Lites, Elmore, \& Streander 2001). 
A SIR inversion of the four Stokes profiles of 2 spectral lines (10
free parameters, 135 wavelength points, model atmosphere discretized
in 41 grid points) takes $0.7\,$s on a dual Xeon workstation running
at $2.8\,$GHz. Optimizing the code, a cadence of $0.5\,$s may easily
be reached. The real-time analysis of a full POLIS slit (450 pixels in
$10\,$s) would then require 20 such workstations.  The analysis of
Solar-B/SP data (1000 pixels every $10\,$s) would require 50
workstations. In both cases, the total cost would be a minor fraction
of the cost of the instruments themselves.

The situation is rather different for vector magnetographs.  These
instruments measure only a few wavelength points, i.e., line profiles
are not available. Probably, the most we can do with this kind of data
is a full least-squares ME inversion.\footnote{However, it remains to
be seen whether vertical gradients can also be recovered from
observations with limited wavelength sampling. Work in this direction
is being done at the IAA in preparation for the analysis of data from 
IMaX and VIM (Orozco Su\'arez, Bellot Rubio, \& del Toro Iniesta 2006).}  
An additional complication is that the data rates will be
huge, much larger than those expected from grating instruments. For
example, HMI will observe about $10^6$ pixels every 80--$120\,$s. To
cope with such data flows, PCA methods and ANNs are being proposed as
the only option to invert the observations in real time.  We have
already mentioned that on average the results of these methods
coincide with those from ME inversions. However, since large errors
occur for many individual pixels, it is clear that ME inversions would
be preferable over PCA or ANN analyses.

But, how to perform ME inversions of vector magnetograph data at the
required speed? The solution could be hardware inversion on Field
Programmable Gate Arrays (FPGAs), which is about 10$^3$ times faster
than software inversion depending on the frequency of the processor
and the implementation of the algorithm. At the IAA, we are studying
the feasibility of such an electronic inversion for the analysis of
VIM data \citep{b3 CL06}.
The first working prototype is expected to be ready by the end of 2007.

\section{Summary}
Inversion techniques (ITs) have become essential tools to investigate the
magnetism of the solar atmosphere. Nowadays, they represent the best
option to extract the information contained in high precision
polarimetric measurements. The reliability and robustness of
least-squares Stokes inversions have been confirmed many times with
the help of numerical tests. Part of the community, however, is still
concerned with uniqueness issues. These concerns will hopefully
disappear with the implementation of more realistic model atmospheres.

During the last years, major progress in the field has resulted from
the application of ITs to state-of-the-art observations. The advent of
spectro-polarimeters for the near IR has represented a
breakthrough, allowing the observation of atomic and molecular lines
that provide increased magnetic and thermal sensitivity, and extended
chromospheric coverage. The potential of simultaneous observations of
visible and IR lines for precise diagnostics of solar magnetic
fields has just started to be exploited. Visible and IR lines
constrain the range of acceptable solutions, which is especially
useful for the investigation of complex structures with different
magnetic components and/or discontinuities along the LOS.
Finally, we have begun to invert spectro-polarimetric
observations at very high spatial resolution, with the aim of reaching
the diffraction limit of current solar telescopes
(0\farcs1--0\farcs2). High spatial resolution allows to separate
different magnetic components that might coexist side by side, thus
facilitating the determination of their properties.

The application of ITs to these observations is casting doubts on the
capabilities of certain lines for investigating particular aspects of
solar magnetism. A detailed study of the limitations of spectral
lines, in particular the often used \ion{Fe}{i} pair at $630\,$nm,
seems necessary to clarify their range of usability. An obvious cure
for any problem that might affect the observables is to invert visible
and IR lines simultaneously. This will require modifications of
current ITs to account for different instrumental effects in the
different spectral ranges.

Perhaps the most important challenge facing ITs in the next years is
the analysis of the enormous data sets expected from upcoming
space-borne polarimeters. So far, the efforts have concentrated on the
development of fast PCA-methods and ANNs for real-time inversions of
the data. However, the unprecedented quality of these observations in
terms of spectral and spatial resolution makes it necessary to explore
the feasibility of more complex inversions capable of determining
gradients of field strength and velocity along the LOS. Tests with
numerical simulations demonstrate the importance of gradients to
reproduce the very large asymmetries of the Stokes profiles expected
at a resolution of 0\farcs1--0\farcs2. Current computational resources
allow us to determine gradients from full line profiles observed with
grating spectro-polarimeters such as the one onboard Solar-B, at a
very reasonable cost. Gradients might also be recovered from
high-resolution filtergraph observations if sufficient wavelength
points are available.  Interestingly, real-time ME inversions of data
with limited wavelength sampling seem possible using FPGAs.  The
feasibility of such electronic ME inversions needs to be assessed. At
the same time, it is important to continue the development of PCA and
ANN methods to provide the more complex ME inversions with good
initial guesses.

\acknowledgements This work has been supported by the Spanish MEC 
through Programa Ram\'on y Cajal and project ESP2003-07735-C04-03.


\begin{thebibliography}{}

\bibitem[Asensio Ramos(2004)]{b3 AR04} 
Asensio Ramos, A. 2004, Ph.D. Thesis, Universidad de La Laguna,
Tenerife, Spain

\bibitem[Asensio Ramos, Trujillo Bueno, \& Collados(2005)]{b3 AR05}
Asensio Ramos, A., Trujillo Bueno, J., \& Collados, M. 2005, \apjl, 623, L57 
 
\bibitem[Auer, House, \& Heasley(1977)]{b3 Au77}
Auer, L. H., House, L. L., \& Heasley, J. N. 1977, \solphys, 55, 47 
 
\bibitem[Beck et al.(2005)]{b3 Be05}
Beck, C., Schmidt, W., Kentischer, T., \& Elmore, D. 2005, \aap, 437, 1159 
 
\bibitem[Beck et al.(2006)]{b3 Be06}
Beck, C., Bellot Rubio, L. R., Schlichenmaier, R., \& S\"utterlin, 
P. 2006, \aap, submitted
 
\bibitem[Bellot Rubio(2003)]{b3 BR03}
Bellot Rubio, L. R. 2003, in ASP Conf. Ser. Vol. 307, Solar 
Polarization 3, ed. J. Trujillo Bueno \& J. S\'anchez Almeida 
(San Francisco: ASP), 301 
 
\bibitem[Bellot Rubio(2004)]{b3 BR04a}
Bellot Rubio, L. R. 2004, Rev.Mod.Astron., 17, 21

\bibitem[Bellot Rubio \& Beck(2005)]{b3 BB05}
Bellot Rubio, L. R., \& Beck, C. 2005, \apj, 626, L125

\bibitem[Bellot Rubio, Balthasar, \& Collados(2004)]{b3 BR04b}
Bellot Rubio, L. R., Balthasar, H., \& Collados, M. 2004, \aap, 427, 319 
 
\bibitem[Be\-llot Rubio, Schlichenmaier, \& Tritschler(2006)]{b3 BR06}
Bellot Rubio, L. R., Schlichenmaier, R., \& Tritschler, A. 2006, 
\aap, 453, 1117
 
\bibitem[Bellot Rubio et al.(2006)]{b3 BR06b} Bellot Rubio, L.~R.,
Tritschler, A., Kentischer, T., Beck, C., \& del Toro Iniesta, J.~C.\
2006, 26th meeting of the IAU, 16-17 August, 2006, Prague, JD03, \#58

\bibitem[Bo\-rrero et al.(2004)]{b3 Bo04}
Borrero, J. M., Solanki, S. K., Bellot Rubio, L. R., Lagg, A., 
\& Mathew, S. K. 2004, \aap, 422, 1093 
  
\bibitem[Bo\-rrero et al.(2005)]{b3 Bo05}
Borrero, J. M., Lagg, A., Solanki, S. K., \& Collados, M. 2005, \aap, 436, 333 

\bibitem[Cabrera Solana, Bellot Rubio, \& del Toro Iniesta(2005)]{b3 CS05}
Cabrera Solana, D., Bellot Rubio, L. R., \& del Toro Iniesta, 
J. C. 2005, \aap, 439, 687 
 
\bibitem[Cabrera Solana et al.(2006)]{b3 CS06}
Cabrera Solana, D., Bellot Rubio, L. R., Beck, C., \& del Toro 
Iniesta, J. C. 2006, these proceedings

\bibitem[Carroll \& Staude(2001)]{b3 CS01}
Carroll, T. A., \& Staude, J. 2001, \aap, 378, 316 
 
\bibitem[Casini et al.(2003)]{b3 Ca03}
Casini, R., L\'opez Ariste, A., Tomczyk, S., \& Lites, B. W. 2003, 
\apjl, 598, L67 
 
\bibitem[Castillo Lorenzo et al.(2006)]{b3 CL06}
Castillo Lorenzo, J. L., Orozco Su\'arez, D., Bellot Rubio, L. R., 
et al.  2006, these proceedings

\bibitem[Cavallini(2006)]{b3 Ca06}
Cavallini, F. 2006, \solphys, 236, 415
 

\bibitem[Frutiger \& Solanki(1998)]{b3 FS98}
Frutiger, C., \& Solanki, S. K. 1998, \aap, 336, L65 
 
\bibitem[Harvey, Livingston, \& Slaughter(1972)]{b3 Ha72}
Harvey, J., Livingston, W., \& Slaughter, C. 1972, in 
Line Formation in the Presence of Magnetic Fields, ed. R. G. Athay, 
L. L. House \& A. Newkirk (Boulder: NCAR),~227 

\bibitem[Kentischer et al.(1998)]{b3 Ke98}
Kentischer, T. J., Schmidt, W., Sigwarth, M., \& Uexk\"ull, M. V. 
1998, \aap, 340, 569 
 
\bibitem[Lagg(2005)]{b3 La05}
Lagg, A. 2005, in ESA SP-596, Proc.Intl.Sci.Conf. Chromospheric 
and Coronal Magnetic Fields, ed. D. E. Innes, A. Lagg, S. K. Solanki
\& D. Danesy (Noordwijk: ESA),~6 

\bibitem[Lagg et al.(2004)]{b3 La04}
Lagg, A., Woch, J., Krupp, N., \& Solanki, S. K. 2004, \aap, 414, 1109 

\bibitem[Lites, Elmore, \& Streander(2001)]{b3 Li01}
Lites, B. W., Elmore, D. F., \& Streander, K. V. 2001, in 
ASP Conf. Ser. Vol. 236, Advanced Solar Polarimetry: Theory, 
Observation, and Instrumentation, ed. M. Sigwarth (San Francisco: ASP), 33 
 
\bibitem[L\'opez Ariste \& Casini(2002)]{b3 LC02}
L\'opez Ariste, A., \& Casini, R. 2002, \apj, 575, 529 
 
\bibitem[L\'opez Ariste \& Casini(2003)]{b3 LC03}
L\'opez Ariste, A., \& Casini, R. 2003, \apjl, 582, L51 
 
\bibitem[L\'opez Ariste \& Casini(2005)]{b3 LC05}
L\'opez Ariste, A., \& Casini, R. 2005, \aap, 436, 325 
 
\bibitem[L\'opez Ariste, Tomczyk, \& Casini(2002)]{b3 LA02}
L\'opez Ariste, A., Tomczyk, S., \& Casini, R. 2002, \apj, 580, 519 
 
\bibitem[L{\'o}pez Ariste et al.(2006)]{b3 LA06} L{\'o}pez 
Ariste, A., Aulanier, G., Schmieder, B., \& Sainz Dalda, A.\ 2006, \aap, 
456, 725 

\bibitem[Mart\'{\i}nez Gonz\'alez, Collados, \& Ruiz Cobo(2006)]{b3 MG06}
Mart\'{\i}nez Gonz\'alez, M. J., Collados, M., \& Ruiz Cobo, B. 
2006, \aap, 456, 1159
 
\bibitem[Mart\'{\i}nez Pillet et al.(1999)]{b3 MP99} 
Mart\'{\i}nez Pillet, V., Collados, M., S\'anchez Almeida, J., et al. 
in ASP Conf. Ser. Vol. 183, High Resolution Solar Physics: 
Theory, Observations, and Techniques, ed. T.~R. Rimmele, K.~S.
Balasubramaniam \& R.~R. Radick (San Francisco: ASP),~264

\bibitem[Mathew et al.(2003)]{b3 Ma03}
Mathew, S. K., Lagg, A., Solanki, S. K., et al. 2003, \aap, 410, 695 
 
\bibitem[Orozco Su\'arez, Lagg, \& Solanki(2005)]{b3 OS05}
Orozco Su\'arez, D., Lagg, A., \& Solanki, S. K. 2005, 
in ESA SP-596, Proc.Intl.Sci.Conf. Chromospheric and Coronal Magnetic 
Fields, ed. D. E. Innes, A. Lagg, S. K. Solanki \& D. Danesy 
(Noordwijk: ESA), 59

\bibitem[Orozco Su\'arez, Bellot Rubio, \& del Toro Iniesta(2006)]{b3 OS06}
Orozco Su\'arez, D., Bellot Rubio, L. R., \& del Toro Iniesta, 
J. C. 2006, these proceedings

\bibitem[Rees et al.(2000)]{b3 Re00}
Rees, D. E., L\'opez Ariste, A., Thatcher, J., \& Semel, M. 2000, 
\aap, 355, 759 
 
\bibitem[Ruiz Cobo \& del Toro Iniesta(1992)]{b3 RT92}
Ruiz Cobo, B., \& del Toro Iniesta, J. C. 1992, \apj, 398, 375 
 
\bibitem[S\'anchez Almeida(2005)]{b3 SA05}
S\'anchez Almeida, J. 2005, \apj, 622, 1292 

\bibitem[Skumanich \& Lites(1987)]{b3 SL87}
Skumanich, A., \& Lites, B. W. 1987, \apj, 322, 473
 
\bibitem[Socas-Navarro(2001)]{b3 SN01a}
Socas-Navarro, H. 2001, in ASP Conf. Ser. Vol. 236, Advanced
Solar Polarimetry: Theory, Observation, and Instrumentation, ed. 
M. Sigwarth (San Francisco: ASP),~487 
 
\bibitem[Socas-Navarro(2003)]{b3 SN03}
Socas-Navarro, H. 2003, Neural Networks, 16, 355 
 
\bibitem[Socas-Navarro(2005a)]{b3 SN05a}
Socas-Navarro, H. 2005a, \apj, 621, 545 
 
\bibitem[Socas-Navarro(2005b)]{b3 SN05b}
Socas-Navarro, H. 2005b, \apjl, 631, L167 

\bibitem[Socas-Navarro, L\'opez Ariste, \& Lites(2001)]{b3 SN01b} 
Socas-Navarro, H., L\'opez Ariste, A., \& Lites, B. W. 2001, \apj, 553, 949 
 
\bibitem[Socas-Navarro et al.(2004)]{b3 SN04}
Socas-Navarro, H., Mart\'{\i}nez Pillet, V., Sobotka, M., \& 
V\'azquez, M. 2004, \apj, 614, 448

\bibitem[Socas-Navarro et al.(2006a)]{b3 SN06a}
Socas-Navarro, H., Elmore, D., Pietarila, A., Darnell, A., 
Lites, B. W., \& Tomczyk, S. 2006a, \solphys, 235, 55 
 
\bibitem[Socas-Navarro et al.(2006b)]{b3 SN06b}
Socas-Navarro, H., Mart\'{\i}nez Pillet, V., Elmore, D., 
Pietarila, A., Lites, B. W., \& Manso Sainz, R. 2006b, \solphys, 235, 75 
 
\bibitem[Solanki et al.(2003)]{b3 So03}
Solanki, S. K., Lagg, A., Woch, J., Krupp, N., \& Collados, M. 2003, 
\nat, 425, 692 
 
\bibitem[Solanki et al.(2006)]{b3 So06}
Solanki, S. K., Lagg, A., Aznar Cuadrado, R., et al. 2006, these proceedings

\bibitem[del Toro Iniesta(2003)]{b3 TI03}
del Toro Iniesta, J. C. 2003, Astron.Nach., 324, 383 
 
\bibitem[del Toro Iniesta \& Ruiz Cobo(1996)]{b3 TR96}
del Toro Iniesta, J. C., \& Ruiz Cobo, B. 1996, \solphys, 164, 169 
 
\bibitem[Trujillo Bueno et al.(2002)]{b3 TB02}
Trujillo Bueno, J., Landi Degl'Innocenti, E., Collados, M., 
Merenda, L., \& Manso Sainz, R. 2002, \nat, 415, 403 
 
\bibitem[V\"ogler et al.(2005)]{b3 Vo05}
V\"ogler, A., Shelyag, S., Sch\"ussler, M., Cattaneo, F., Emonet, T., 
\& Linde, T.  2005, \aap, 429, 335 
 
\bibitem[Westendorp Plaza et al.(2001)]{b3 WP01}
Westendorp Plaza, C., del Toro Iniesta, J. C., Ruiz Cobo, B., 
Mart\'{\i}nez Pillet, V., Lites, B. W., \& Skumanich, A. 2001, \apj, 547, 1130 
 
\end{thebibliography}
\end{document}